\documentclass{article}

\usepackage{fullpage}
\usepackage{natbib}
\usepackage{amsmath,amsfonts}
\usepackage{algorithm,algorithmic}
\usepackage{amsthm}
\newtheorem{definition}{Definition}
\newtheorem{result}{Result}
\usepackage{mathtools}     % an extension package to amsmath
\mathtoolsset{showonlyrefs}
\usepackage{setspace}                  % Deux lignes pour un espace interligne double
\usepackage{multirow}
\usepackage{booktabs}
\usepackage{tikz}

\def\cb{\mathbf{c}}
\def\pb{\mathbf{p}}

\def\sb{\mathbf{s}}
\def\ub{\mathbf{u}}
\def\xb{\mathbf{x}}
\def\zb{\mathbf{z}}
\def\Ab{\mathbf{A}}
\def\Sb{\mathbf{S}}
\def\0b{\mathbf{0}}
\def\1b{\mathbf{1}}

\def\R{\mathbb{R}}

\def\pib{\boldsymbol{\pi}}

\def\RCmean{\textup{RC}_{\textup{mean}}}
\def\RCmin{\textup{RC}_{\textup{min}}}
\def\RCmax{\textup{RC}_{\textup{max}}}

\def\Cmean{\widetilde{c}_{\textup{mean}}}
\def\Cmin{\widetilde{c}_{\textup{min}}}
\def\Cmax{\widetilde{c}_{\textup{max}}}

\title{In Search of the Most Balanced Sampling Design}

\author{Caren Hasler\thanks{Department of Psychology,
    Psychological Methods, Evaluation and Statistics,
    University of Zurich, Switzerland. Email: caren.hasler\text{@}psychologie.uzh.ch},
    Esther Eustache\thanks{Institute of Sport Sciences, University of Lausanne, Switzerland. Email: esther.eustache\text{@}unil.ch},
        Yves Till\'e\thanks{University of Neuch\^atel, Switzerland. Email: yves.tille\text{@}unine.ch}}

\begin{document}
%--------------------------------------------------------

\maketitle

\noindent\textbf{Data Availability Statement.}
The simulation data used in this study can be reproduced using the settings described in the manuscript. The real data set (MU284) is publicly available in the \texttt{sampling} R package \citep{til:mat:25}.

\vspace{0.5em}
\noindent\textbf{Conflict of Interest Statement.}
The authors declare that they have no conflicts of interest.

\vspace{0.5em}
\noindent\textbf{Funding Statement.}
This research received no specific grant from any funding agency.

\vspace{0.5em}
\noindent\textbf{Statement of Artificial Intelligence Use.}
Artificial intelligence (AI) tools, including ChatGPT (OpenAI), Claude (Anthropic), and DeepL, were used to improve language, readability, and clarity and to assist in drafting selected code and figures. All AI-generated outputs were reviewed, verified, and edited by the authors. The authors are solely responsible for the scientific content presented in this work.

\begin{abstract}
Balanced sampling aims to select random samples in which the estimated totals of the auxiliary variables, weighted by the inverse of the inclusion probabilities, correspond as closely as possible to the known population totals. 
While several methods, such as rejective sampling, rerandomization, and the cube method, have been proposed to improve balance, identifying the most balanced sampling design under fixed inclusion probabilities remains a challenging combinatorial problem. This problem can be formulated as a linear program defined over the set of all possible samples, but the number of samples grows exponentially with population size, making exact optimization infeasible except for very small populations. 
To address this issue, we propose a heuristic approach based on a genetic algorithm that iteratively improves the balance of sampling designs by combining minimum support designs with highly balanced candidate samples. 
Although optimality cannot be guaranteed, the proposed method can substantially improve balance relative to standard procedures such as the cube method. 
The approach is applicable to both survey sampling and experimental design.

\end{abstract}

%--------------------------------------------------------
\section{Introduction}
%--------------------------------------------------------

The problem of balanced sample selection has been studied since the early development of sampling theory. \citet{gin:gal:29} had already selected a sample of Italian districts that was balanced across several demographic variables. 
The sample was not selected randomly, and the balance was achieved through trial and error. 
\citet{yat:49} and \citet{thi:53} then proposed iterative methods based on unit exchanges. 
\citet{haj:64,haj:81} subsequently developed a theory of rejective sampling, which consists of selecting a sequence of samples until a sufficiently balanced sample is obtained \citep[see also][]{dup:79,fuller2009some,boistard2012approximation,fuller2017bootstrap}.

Rerandomization is the analog of rejective sampling in the field of experimental design and has recently attracted renewed interest
\citep[see for instance][]{lock2011,rubin2012,morgan2015rerandomization,zhou2018sequential,li2018asymptotic,li2020rerandomization,Tille2021,KAPELNER2022,yang2023rejective}. With rerandomization, units are repeatedly assigned at random to treatment groups until a balance criterion on the observed covariates is met. Creating treatment groups in experimental designs is a balanced sample selection problem: one seeks to select samples (the treatment groups) as balanced as possible on the observed covariates. In this context, the inclusion probabilities correspond to treatment assignment probabilities. When partitioning into treatment and control groups of equal size, rerandomization is a special case of rejective sampling where the inclusion probabilities are all equal to one-half \citep{Tille2021}.

Even though the aforementioned methods allow us to select balanced samples, they do not necessarily preserve exactly the predetermined inclusion probabilities. 
Indeed, excluding some samples from the sampling design or, equivalently, some group assignments, modifies the inclusion probabilities. 
These probabilities can nevertheless be estimated \citep{cha:haz:les:17}.

When one seeks to exactly respect predetermined inclusion probabilities, a more appropriate method is the cube method of \citet{dev:til:04a}. This method randomly selects a balanced or approximately balanced sample while exactly respecting predetermined inclusion probabilities. The cube method was developed further and adapted in the context of sampling \citep[see for instance][]{cha:til:06,cha:09,brei:chauv:12,has:til:14,eus:jaus:til:2021,leu:eus:jau:til:2022} and rerandomization \citep{Tille2021,ejub2025risk,dav:al:25}. This method is easy to implement but does not necessarily provide the most balanced sampling design.

If a balancing criterion is available for each sample, the search for the most balanced design can be written as a linear program. This approach was used by \citet{ard:91} to select the primary units of the French master sample. \citet{dev:til:04a} use the same method for the landing phase of the cube method. However, this approach cannot be applied to populations of more than 20 units because the number of samples to be considered increases exponentially with the size of the population. Its field of application is therefore extremely limited. 
Another alternative was proposed by \cite{benedetti2022simulated} who introduced a simulated annealing approach to select fixed-size samples that minimize a balancing criterion. 
While their method provides promising results in terms of balancing, it does not respect predetermined inclusion probabilities.

The genetic algorithm developed by \cite{hol:75} is an optimization technique inspired by the concept of natural evolution proposed by Charles Darwin. 
It involves repeatedly modifying a population of solutions to an optimization problem. 
At each iteration, the refined offspring are considered as the next-generation population of solutions. 
To the best of our knowledge, this approach has been used in the context of survey sampling to jointly determine optimal stratification and sample allocation by \cite{BallBarc2013,Barsamp2014,olu:al:19} and was extended to spatial sampling by \cite{bal:20}.

This article presents a novel method for constructing balanced sampling designs using a genetic algorithm. Our approach iteratively improves the balance of a set of designs at each iteration. While we cannot guarantee identification of the globally optimal balanced design, the improvement in balance can be substantial compared to the cube method. This method thus has the potential to improve the balance in practical applications in both sampling and experimental designs.

The paper is organized as follows. We introduce the problem and some pieces of notation in Section~\ref{sec:problemandnotation}. The proposed algorithm is presented in Section~\ref{sec:method}. Section~\ref{sec:simu} contains the results of simulation studies. The main messages of the current work are summarized in Section~\ref{sec:conclusion}. 
%The Appendices contain some technical elements. 

%--------------------------------------------------------
\section{Problem and Notation} \label{sec:problemandnotation}
%--------------------------------------------------------

Consider a finite population $U=\{1,\dots,k,\dots,N\}$ from which we wish to randomly select a sample of fixed size $n$. A sample is a subset of the population, represented by a vector $\sb=(s_1,\dots,s_k,\dots,s_N)^\top$ where $s_k$ is 1 if unit $k$ is selected and 0 otherwise. Let $\pib=(\pi_1,\dots,\pi_k,\dots,\pi_N)^\top$ denote the vector of prescribed first-order inclusion probabilities, defined prior to sample selection. The objective is to select a random sample such that each unit $k$ is included with probability $\pi_k$. 

There are
\begin{equation}
 M=\binom{N}{n}=\frac{N!}{n!(N-n)!}
\end{equation}
possible samples of size $n$ in a population of size $N$. Let $\Sb$ denote the matrix of size $N\times M$ whose columns contain the $M$ possible samples. A sampling design is a probability distribution over all possible samples, or, equivalently over the columns of $\Sb$. Hence, for a given matrix $\Sb$, a sampling design is represented by a vector of sample selection probabilities $\pb=(p_1,\dots,p_i,\dots,p_M)^\top$ such that
\begin{equation}
p_i \ge 0,~ i=1,\dots, M, \mbox{ and } \sum_{i=1}^{M}p_i =1.
\end{equation}
A sampling design satisfies the inclusion probabilities if %the corresponding matrix of samples $\Sb$ and vector of selection probabilities $\pb$ satisfy
$\Sb\;
\pb = \pib.$

Now suppose that the values of $p$ auxiliary variables are known for all units of the population. Let $\xb_k=(x_{k1},\dots,x_{kp})^\top$ denote the vector of values taken by these $p$ auxiliary variables for unit $k$. A sample is perfectly balanced on these variables if their Horvitz-Thompson estimators are equal to their population totals, in other words, if
\begin{equation}\label{eqn:balancing}
\sum_{k\in U} \frac{\xb_k s_k}{\pi_k} = \sum_{k\in U} \xb_k.
\end{equation}
It is often impossible to obtain a sample that is exactly balanced. Indeed, a sample is a vector containing only zeros and ones, and it may be that no such vector exactly satisfies the balancing Equation~\eqref{eqn:balancing}.

The quality of the balance of a sample $\sb$ can be assessed using a cost function $c(.)$, which quantifies the discrepancy between the Horvitz-Thompson estimators of the auxiliary variables and their population totals. The better the balance, the smaller the cost. Several cost functions can be considered. \citet{dev:til:04a} suggest
\begin{equation}
c(\sb) = (\sb - \pib)^\top \Ab^{\top} \left( \Ab \Ab^\top\right)^{-1} \Ab (\sb-\pib),
\end{equation}
where $\Ab$ is a $p\times N$ matrix defined by
\begin{equation}
\Ab =\left(\xb_1/\pi_1,\dots,\xb_k/\pi_k,\dots,\xb_N/\pi_N\right).
\end{equation}
This cost $c(\sb)$ is invariant under scaling (i.e., multiplication by a constant) of the auxiliary variables and $c(\sb)=0$ if sample $\sb$ is perfectly balanced. We consider this cost function throughout the manuscript. However, our results remain valid for other choices.
% For \violet{instance}, one may consider the function \textcolor{red}{XXXEst-ce nécessaire de mettre cela ?XXX}
% \begin{equation}
% c(\sb) = \sum_{j=1}^p \frac{1}{V_j^2} \left(\sum_{k\in U} \frac{x_{kj} s_k}{\pi_k} - \sum_{k\in U} x_{kj}\right)^2,
% \end{equation}
% where
% \begin{equation}
% V^2_j = \frac{1}{N}\sum_{k\in U}(x_{kj}-\overline{X}_j)^2
% \end{equation}
% and
% \begin{equation}
% \overline{X}_j = \frac{1}{N}\sum_{k\in U} x_{kj}.
% \end{equation} 
For a given cost function, we define the vector $\cb$ of dimension $M$, which contains the cost of the possible samples, i.e., of the columns of $\Sb$. 
The cost of a sampling design is then defined as the average cost of the samples weighted by their respective selection probabilities, i.e., $\tilde{c} = \pb^\top\cb.$

The optimal design, in the sense of being the most balanced, is the sampling design with minimum cost among the sampling designs that satisfy the inclusion probabilities. Mathematically, the optimal design is the solution to the following linear program
\begin{equation}
\min_{\pb \in \left[0,1 \right]^M} \tilde{c} \label{eq:proglin}
\end{equation}
subject to
\begin{equation}
\Sb \; \pb = \pib \;\; \mbox{ and }\;\;\1b^\top \pb=1,
\end{equation}
where $\1b$ is a column vector of dimension $M$ containing only ones. 
This program can be solved using an algorithm for linear programming such as the simplex algorithm. 
However, the method cannot be applied to populations larger than approximately 30 units, because the number of samples to be considered increases exponentially with the size of the population. 
In this article, we propose an algorithm for obtaining a sampling design that approaches the optimal design as closely as possible while remaining applicable to large populations.

%--------------------------------------------------------
\section{A Genetic Algorithm to Search the Most Balanced Sampling Design}\label{sec:method}
%--------------------------------------------------------

Two building blocks of the proposed algorithm are minimum support designs and highly balanced samples. 
Procedures to generate highly balanced samples are described in Section~\ref{sec:highly:balanced} and minimum support designs are discussed in Section~\ref{sec:min:support}. Two versions of the proposed algorithm are presented in Sections~\ref{sec:algorithm1} and \ref{sec:algorithm2}, respectively. %Variance estimation is discussed in Section~\ref{sec:var}. 
Finally, we present some aborted alternatives that have shown to be less effective than the proposed algorithms in Section~\ref{sec:abrtd}.

%--------------------------------------------------------
\subsection{Generating Highly Balanced Samples}\label{sec:highly:balanced}

Consider the polytope
\begin{equation}
K = \left\{\ub \in \R^N \mid \ub\in [0,1]^N \mbox{ and } \Ab\ub=\Ab \pib \right\}.
\end{equation}
By construction, any vector $\ub$ in $K$ satisfies the balancing Equation~\eqref{eqn:balancing}. %because
%\begin{equation}
%\Ab\ub &= \sum_{k\in U} \frac{\xb_k u_k}{\pi_k}, \\
%\Ab \pib &= \sum_{k\in U} \xb_k.
%\end{equation}
To select highly balanced samples, a vertex $\pib^*$ of the polytope $K$ is first selected. This can be achieved in several ways, including the following two approaches.
%\begin{itemize}
%\item

The first approach is to apply the flight phase of the cube method proposed by \citet{dev:til:04a}, which randomly selects a vertex $\pib^*$ of $K$ such that $\mathrm{E}_p(\pib^*) = \pib$, where $\mathrm{E}_p$ denotes the expectation with respect to the sampling design.
%\item
The second approach is proposed in \citet{tilleab2026} and consists of first generating a vector $\zb$ containing $N$ independent standard normal random variables. Then, the following linear program is solved
\begin{equation}
 \max_{\ub \in K} \ub^\top \zb,
\end{equation}
to which the solution is a vertex of $K$.
%\end{itemize}

Irrespective of the chosen approach, the selected vertex exactly satisfies the balancing equations. However, a vertex is in principle not a sample, since not all of its components are necessarily equal to 0 or 1. The following result, given in \citet{dev:til:04a}, clarifies this point.
\begin{result}
The vertices of the polytope $K$ contain at most $p$ components that are neither 0 nor 1, where $p$ is the number of rows of $\Ab.$
\end{result}
Since $\Ab$ has $p$ rows, $N-p$ of the components of the vertex $\pib^*$ are either 0 or 1, but the remaining $p$ components may take values strictly between 0 and 1. The selected vertex must therefore be converted into a highly balanced sample.
To do so, a solution is to enumerate all samples $\sb$ of size $n$ having the same integer components as $\pib^*$, and select the one with lowest cost $c(\sb)$.

%i.e., $s_k = \pi_k^*$ when $\pi_k^*\in \{0,1\}$.
%\violet{The following procedure} describes how to convert a selected vertex into a highly balanced sample:
%\begin{itemize}
%\item We randomly select a vertex of $K$, denoted $\pib^*$, using one of the two aforementioned options.
%\item We list all possible samples $\sb$ of size $n$ having the same integer components as $\pib^*$, i.e., $s_k = \pi_k^*$ when $\pi_k^*\in \{0,1\}$. We select the sample with the lowest cost $c(\sb)$.
%\end{itemize}

%--------------------------------------------------------
\subsection{Mininum Support Designs and Systematic Sampling}\label{sec:min:support}
In this section, we introduce the notion of support of a sampling design and recall some important properties of minimum support designs. 
These concepts are key to the performance of the proposed algorithms.

\begin{definition}
The support $\mathcal{Q}$ of a sampling design $\pb$ on $\Sb$ is the set of samples in $\Sb$ that have a non-zero probability of being selected.
\end{definition}

\begin{definition}
A sampling design $\pb$ on $\Sb$ with inclusion probabilities $\pib = \Sb \; \pb$ and support $\mathcal{Q}$ has minimum support if there exists no proper subset of $\mathcal{Q}$ on which a sampling design with inclusion probabilities $\pib$ can be defined.
\end{definition}

\begin{result}\label{result:card:min:support}
The cardinality of the support of a minimum support design is less than or equal to the population size~$N$.
\end{result}
\noindent The proof is given in \citet{wyn:77} using Carathéodory's theorem \citep[see also][]{pea:qua:til:07}. Minimum support designs are one of the building blocks of our proposed algorithms. To obtain such designs, we exploit a result shown in \citet{pea:qua:til:07} and stating that a systematic sampling design with unequal inclusion probabilities has minimum support. Systematic sampling design with unequal inclusion probabilities was proposed by \citet{mad:49} \citep[see also][]{bre:han:83,til:06}. 

A procedure to obtain systematic sampling design is presented in what follows. This procedure requires to select a systematic sample. We first describe a procedure of systematic sample selection.
First, the cumulative inclusion probabilities
$$
V_k = \sum_{j=1}^k \pi_j,
$$
are computed for $k=1,\dots,N$ with $V_0=0$ and $V_N=n$. 
Then, if no starting value $u$ is specified, one is randomly generated by drawing from a continuous uniform random variable on $[0,1]$. 
Finally, the $n$ units $k_1,\dots,k_j,\dots,k_n$ are selected such that $u+j,j=0,\ldots,n$ falls between two values $V_{k_j-1}$ and $V_{k_j}$.
The corresponding algorithm is provided in Appendix~\ref{appendix:systematicsample}.

%\begin{algorithm}[h]
%\caption{Algorithm for systematic design with unequal inclusion probabilities}\label{algo:systematic:sample}
%\textbf{Input}: vector of inclusion probabilities $\pib$ of size $N$ and starting value $u\in [0,1]$ (optional).\\
%\textbf{Output}: Sample $\sb$.
%\begin{itemize}
%\item Compute the cumulative inclusion probabilities $V_k = \sum_{j=1}^k \pi_j$ for $k=1,\dots,N$. Set $V_0=0$ and $V_N=n$. \orng{If sum of pi is integer, then no need to define VN. Should we focus on the case where this sum is integer?}
%\item If no value is specified for the starting value $u$, generate a value $u$ from a random draw of a continuous uniform random variable defined on the interval $[0,1]$.
%\item Define the vector $\sb$ of length $N$ with all elements equal to 0.
%\item For $j = 1, \ldots, n$
%    \begin{itemize}
%        \item Set $s_k = 1$ for $k$ such that $V_{k-1} < u + (j-1) \leq V_{k}, k = 1, \ldots N$.
%    \end{itemize}
%\item Return the sample $\sb$.
%\end{itemize}
%\end{algorithm}

We now describe how to build a systematic design with unequal inclusion probabilities. The corresponding algorithm, described in Appendix~\ref{appendix:systematicdesign}, follows from Results~1 and~2 in \citet{pea:qua:til:07} and is implemented in the \texttt{UPsystematicpi2} function of the \texttt{R} package \texttt{sampling} \citep{til:mat:25}. %This function is designed to compute the joint inclusion probabilities of a systematic design.

First, we compute the cumulative inclusion probabilities $V_k$ as described above and define $v_k=(V_k \bmod 1)$ as the fractional part of $V_k$. Then we sort the fractional parts to obtain $v_{(k)}, k=0,\dots,N-1$ with $v_{(0)}=0$. We set $v_{(N)}=1$. 
The sequence of values $v_{(k)}, k=0,\dots,N$ creates a partition of the interval $[0,1]$ such that the algorithm returns the same sample for any starting value $u$ between two consecutive values in $v_{(k)},k=0,\dots,N$.

For example, if we consider any starting value $u$ such that $v_{(0)} < u \leq v_{(1)}$, then we select the same sample, say $\sb_1$. Moreover, if we consider any starting value $u$ such that $v_{(1)} < u \leq v_{(2)}$, then we select the same sample, say $\sb_2$. However, samples $\sb_1$ and $\sb_2$ are different. This continues until $v_{(N-1)}$ and $v_{(N)}$. Then, we obtain midpoints $r_k = \left(v_{(k)} - v_{(k-1)} \right)/2, k = 1, \dots, N$ between the consecutive values in $v_{(k)},k=0,\dots,N$. The values $r_k, k = 1, \dots, N$ are representative starting values that are later used to select systematic samples. We compute the corresponding probabilities of selection $p_k=v_{(k)}-v_{(k-1)},k=1,\dots,N$, which correspond to the lengths of the interval defined by the consecutive values in $v_{(k)},k=0,\dots,N$. Finally, for every non-zero probability of selection $p_k$, we select the systematic sample with corresponding starting value $r_k$ as described earlier in the current Section. Putting all the selected samples as columns of a matrix $\Sb$ and the corresponding probabilities of selection in a vector $\pb$, we obtain a systematic sampling design $\pb$ on $\Sb$ that satisfies the inclusion probabilities.

%\begin{algorithm}[h]
%\caption{Algorithm for systematic design with unequal inclusion probabilities}\label{algo:systematic:design}
%\textbf{Input}: vector of inclusion probabilities $\pib$ of size $N$ with integer sum $\sum_{k \in U} \pi_k$.\\
%\textbf{Output}: Sampling design $\pb$ on matrix $\Sb$.
%\begin{itemize}
%\item Compute the cumulative inclusion probabilities $V_k = \sum_{j=1}^k \pi_j,k = 1, \ldots, N$.
%\item Compute $v_k=(V_k \bmod 1)$ the fractional part of $V_k$.
%\item Build $v_{(k)}$ the ordered $v_k$ for $k=0,\dots,N$. Set $v_{(N+1)}=1$.
%\item Compute $r_k =(v_{(k)} + v_{(k+1)})/2$ for $k=0,\dots,N$.
%\item Compute $p_k = v_{(k+1)} - v_{(k)}$ for $k=0,\dots,N$.
%\item Remove from $\pb$ any null element. Remove the corresponding elements from $\rb$.
%\item Repeat for $k = 1, \ldots, \mbox{length}(\pb)$
%\begin{itemize}
% \item Select the systematic sample with inclusion probabilities $\pib$ and starting value $u = r_k$ with Algorithm~\ref{algo:systematic:sample}.
% \item The selected sample is the $k$-th column of $\Sb$.
%\end{itemize}
%\item Return $\pb$ and $\Sb$.
%\end{itemize}
%\end{algorithm}

It is therefore possible to generate a systematic design with unequal inclusion probabilities in just a few lines of code. Since the systematic design depends on the order of the population units and is a minimum support design, each non-circular permutation of these units provides another minimum support design.

Another way to generate minimum support designs is to use the minimum support procedure described in \citet{dev:til:98}. This procedure is a special case of the splitting method. It has the advantage of being able to force the presence of a particular sample in the design. Finally, an important result is given in \citet{dev:til:04a}.
\begin{result}\label{result:sol:lin:prog}
The linear program given in Expression~\eqref{eq:proglin} has at least one solution that is a minimum support design.
\end{result}
%Since the simplex algorithm terminates at a vertex of the simplex, it will necessarily terminate on a minimum support design.
Results~\ref{result:card:min:support} and \ref{result:sol:lin:prog} jointly imply that the optimal design contains at most $N$ samples with a non-null probability of selection. This is central to the success of our proposed algorithm described in the section that follows.

\subsection{Algorithm~1}\label{sec:algorithm1}
%--------------------------------------------------------

The first version of the method is presented in Algorithm~\ref{algo:yves} and detailed here. 
First, $R \in \mathbb{N}$ random permutations of the population are generated and the corresponding systematic sampling designs are obtained. See Section~\ref{sec:min:support} for more details. The generated designs have minimum support \citep{pea:qua:til:07} and contain at most $N$ samples with non-zero selection probabilities (see Result~\ref{result:card:min:support}). The matrix whose columns are the samples in the support of the design and corresponding vectors of selection probabilities and sample costs are constructed for each design.

Next, the initially generated designs are iteratively improved by reducing their cost as follows.
At each iteration, two designs among the $R$ candidate designs are selected at random. 
Then, $D$ well-balanced samples are generated as explained in Section~\ref{sec:highly:balanced}. 
The linear program in Expression~\eqref{eq:proglin} is then applied to the restricted set of samples consisting of those from the two selected designs and the $D$ well-balanced samples. 
This set contains at most $2N + D$ samples. 

The solution of the linear program defines a new sampling design whose support contains at most $N$ samples with non-zero selection probabilities, as stated in Result~\ref{result:sol:lin:prog}. 
This new design replaces the design with the largest cost among the $R$ candidate designs. 
At the end of each iteration, exactly one design has been replaced, while the remaining $R-1$ designs are unchanged. 

The iterations are repeated with the updated set of $R$ candidate designs until a stopping criterion is reached.
Several stopping criteria may be considered. 
For example, the iterations may be terminated after a fixed number of iterations, after a given computation time, when the cost of the designs falls below a target value, or when the relative difference in mean cost of the designs between two consecutive iterations falls below a target value.
At the end of the iterations, we obtain $R$ well-balanced designs. 
This procedure can be interpreted as a genetic algorithm in which two designs are combined and improved through the addition of highly balanced samples.

At the end of Algorithms~\ref{algo:yves}, we have $R$ well-balanced sampling designs. 
Several options are available to obtain a single well-balanced design from these $R$ designs. 
For example, one may consider the design with the lowest cost among these designs. 
One may also combine these $R$ designs pairwise using the linear program until only one remains. 

\begin{algorithm}[htb!]
\caption{Genetic algorithm for balanced sampling 1}\label{algo:yves}
\begin{algorithmic}[1]
\REQUIRE Inclusion probabilities $\pib = ( \pi_1,\dots,\pi_N)^\top$ with $\sum_{k \in U}\pi_k = n$; number of candidate designs $R$;
number of added highly balanced samples $D$; stopping criterion $\mathcal{C}$
\ENSURE A set of $R$ well-balanced designs $\{\mathcal{P}_1,\dots,\mathcal{P}_R\}$

\STATE \textbf{Initialization}
\FOR{$r=1$ \TO $R$}
    \STATE $\mathcal{P}_r \leftarrow$ random systematic design satisfying $\pib$
    %\COMMENT{$\mathcal{P}_r$ has \vert{minimum} support with cardinal $\leq N$}
\ENDFOR

\STATE \textbf{Iterative improvement}
\WHILE{$\mathcal{C}$ not fulfilled}
    \STATE Randomly select two distinct indices $a,b \in \{1,\dots,R\}$
    \STATE Generate $\mathcal{B} \leftarrow$ a set of $D$ highly balanced samples (Section~\ref{sec:highly:balanced})
    \STATE $\mathcal{S} \leftarrow support (\mathcal{P}_a)\ \cup\ support(\mathcal{P}_b) \cup \mathcal{B}$
    %\COMMENT{$|\mathcal{S}|\le 2N+D$}

    \STATE $\mathcal{P}_{\text{new}} \leftarrow$ solve linear program in Expression~\eqref{eq:proglin} with $\mathcal{S}$
    %\COMMENT{$|support(\mathcal{P}_{\text{new}})|\le N$}

    \STATE $r^\star \leftarrow \arg\max_{r\in\{1,\dots,R\}} \tilde{c}(\mathcal{P}_r)$
    \STATE $\mathcal{P}_{r^\star} \leftarrow \mathcal{P}_{\text{new}}$
    %\COMMENT{Replace the currently most costly design}
\ENDWHILE

\RETURN $\{\mathcal{P}_1,\dots,\mathcal{P}_R\}$
\end{algorithmic}
\end{algorithm}

\subsection{Algorithm~2}\label{sec:algorithm2}

The second proposed approach is presented in Algorithm~\ref{algo:esther:caren} and detailed here. First, $R \in \mathbb{N}$ random permutations of the population are generated and the corresponding systematic sampling designs are obtained as explained in Section~\ref{sec:min:support}. In each iteration, $R$ sets of samples are formed by randomly selecting couples or triplets of set of samples from $R+1$ available sets of samples made of the support of the $R$ candidate sampling designs, together with the set of $D$ well-balanced samples generated as explained in Section~\ref{sec:highly:balanced}. Thus, the pairs or triplets are selected from the $R+1$ available sets of samples. The linear program in Expression~\eqref{eq:proglin} is then applied to each of the $R$ pairs or triplets of sets of samples. All $R$ input sampling designs are replaced with the $R$ solutions to the linear program. The iterations are repeated until a stopping criterion is reached, see Section~\ref{sec:algorithm1}. At the end of Algorithm~\ref{algo:esther:caren}, $R$ well-balanced sampling designs are obtained. For details on how to select one of these designs, see Section~\ref{sec:algorithm1}.

\begin{algorithm}[htb!]
    \caption{Genetic algorithm for balanced sampling 2}\label{algo:esther:caren}
    \begin{algorithmic}[1]
        \REQUIRE Inclusion probabilities $\pib = ( \pi_1,\dots,\pi_N)^\top$ with $\sum_{k \in U}\pi_k = n$; number of candidate designs $R$;
        number of added highly balanced samples $D$; stopping criterion $\mathcal{C}$
        \ENSURE A set of $R$ well-balanced designs $\{\mathcal{P}_1,\dots,\mathcal{P}_R\}$

        \STATE \textbf{Initialization}
        \FOR{$r=1$ \TO $R$}
            \STATE $\mathcal{P}_r \leftarrow$ random systematic design satisfying $\pib$
            %\COMMENT{$\mathcal{P}_r$ has \vert{minimum} support with cardinal $\leq N$}
        \ENDFOR

        \STATE \textbf{Iterative improvement}
        \WHILE{$\mathcal{C}$ not fulfilled}
            \STATE Generate $\mathcal{B} \leftarrow$ a set of $D$ highly balanced samples (Section~\ref{sec:highly:balanced})
            \STATE Define the $R+1$ available sample-sets:
            \STATE \hspace{1em} $\mathcal{Q}_0 \leftarrow \mathcal{B}$ and $\mathcal{Q}_r \leftarrow support (\mathcal{P}_r)$ for $r=1,\dots,R$

            %\STATE Initialize an empty list of new designs $\{\mathcal{P}^{\text{new}}_1,\dots,\mathcal{P}^{\text{new}}_R\}$
            \FOR{$r=1$ \TO $R$}
                \STATE Randomly select either a \textbf{pair} or a \textbf{triplet} of distinct indices
                \STATE \hspace{1em} $I_r \subset \{0,1,\dots,R\}$ with $|I_r|\in\{2,3\}$
                \STATE Construct the restricted sample set
                \STATE \hspace{1em} $\mathcal{S}_r \leftarrow \bigcup_{i\in I_r}\mathcal{Q}_i$
                \STATE Solve linear program given in Expression~\eqref{eq:proglin} on $\mathcal{S}_r$ to obtain
                \STATE \hspace{1em} $\mathcal{P}^{\text{new}}_r \leftarrow$ linear program solution design on $\mathcal{S}_r$
                %\COMMENT{$|\supp(\mathcal{P}^{\text{new}}_r)|\le N$}
            \ENDFOR

            \STATE Replace all designs simultaneously:
            \STATE \hspace{1em} $\mathcal{P}_r \leftarrow \mathcal{P}^{\text{new}}_r$ for all $r=1,\dots,R$
        \ENDWHILE

        \RETURN $\{\mathcal{P}_1,\dots,\mathcal{P}_R\}$
    \end{algorithmic}
\end{algorithm}

Algorithms~\ref{algo:yves} and~\ref{algo:esther:caren} differ in their iterations. 
In each iteration of Algorithm~\ref{algo:yves}, only one linear program is solved and a single sampling design is updated. 
By contrast, in Algorithm~\ref{algo:esther:caren}, $R$ linear programs are solved and all $R$ sampling designs are updated. 
Thus, each iteration is more computationally costly in the second algorithm, but the balance increases faster in one iteration. 
However, if both algorithms are run for the same amount of time, then the final sampling designs achieve a similar level of balance. 
Algorithm~\ref{algo:esther:caren} could be sped up by running the $R$ linear programs in parallel. However, no such parallelization is available for Algorithm~\ref{algo:yves}. 
Hence, while both algorithms achieve comparable levels of balance for small to medium populations, Algorithm~\ref{algo:esther:caren}, when implemented in parallel, attains a higher level of balance for large populations within a given time.

\subsection{Aborted Alternatives}\label{sec:abrtd}
%--------------------------------------------------------

We have tested two alternatives to the presented algorithms that have shown to be ineffective. We briefly describe these alternatives below. The first alternative consists in generating a large number of well-balanced samples as described in Section~\ref{sec:highly:balanced} and applying the linear program in Expression~\eqref{eq:proglin} only to these samples. Although this solution is easy to implement, it is unfortunately ineffective. Indeed, a very large number of well-balanced samples must be generated for the linear program to admit a solution and, even with a very large number of samples, there is no guarantee that a solution exists. Hence, in most cases, this alternative fails at generating balanced sampling designs that satisfy the inclusion probabilities.

The second alternative consists in replacing the $D$ highly balanced samples in Algorithms~\ref{algo:yves} and~\ref{algo:esther:caren} by a systematic sampling design. We thought this could improve the entropy of the final sampling designs. When the procedures are iterated for the same amount of time, the final sampling designs obtained with this alternative have a lower level of balance (i.e., a higher cost) than those obtained with Algorithms~\ref{algo:yves} and~\ref{algo:esther:caren}, without increasing the entropy. Therefore, this alternative shows no advantage as compared to the proposed algorithms.

%--------------------------------------------------------
\section{Simulation Study}\label{sec:simu}
%--------------------------------------------------------

\subsection{Simulated data} \label{sec:simu:simulateddata}

To evaluate the proposed algorithms under various configurations, two finite populations of sizes $N=200$ and $N=500$ are generated. 
Each population includes auxiliary variables $X_1, \dots, X_{15}$, generated from a multivariate normal distribution with marginal mean of $50$ and uniform pairwise correlation of $0.4$.
The inclusion probabilities $\pib = (\pi_1, \dots, \pi_N)$ are set proportional to $X_1$ and such that the sample size is $n = \sum_{k=1}^{N}\pi_k = N/4$.
The goal is to obtain a sampling design satisfying the inclusion probabilities~$\pib$, while achieving the best possible balance on a set of auxiliary variables. We consider two settings in which the balancing variables are $X_1, \dots, X_p$, with $p = 10$ and $p=15$, respectively.

For each combination of population size ($N=200, 500$) and number of balancing variables ($p = 10, 15$), we generate $R$ well-balanced designs using Algorithms~\ref{algo:yves} and ~\ref{algo:esther:caren}.
For both algorithms, the number of candidate designs is set to $R=8$, the number of added highly balanced samples to $D=1$, generated using the first approach detailed in Section~\ref{sec:highly:balanced}, and the stopping criteria $C$ is defined as when the difference in mean cost of the $R$ candidate designs between two consecutive iterations falls below $\varepsilon = 10^{-4}$.

To evaluate the proposed methods, we compare the balance they achieve against that of the cube method \citep{dev:04}, which is, to the best of our knowledge, the only available approach to select balanced samples while satisfying inclusion probabilities.
However, since the cube method does not yield a closed-form expression for its sampling design, the cost $\widetilde{c}$ cannot be evaluated analytically. 
We therefore estimate it via Monte Carlo simulation: for each configuration, $I = 10{,}000$ balanced samples $\sb^{(i)}$, $i = 1, \dots, I$, are independently generated using the cube method under the inclusion probabilities $\pib$, and the cost is estimated by
\begin{equation}
    \widehat{\widetilde{c}}_{\,\textup{cube}} = \frac{1}{I} \sum_{i=1}^{I} c\!\left(\sb^{(i)}\right).
    \label{eq:cost_cube}
\end{equation}
We use the function \texttt{cube} from the \texttt{BalancedSampling} package with the default settings to obtain each balanced sample $\sb^{(i)}$.

The two proposed methods produce $R$ well-balanced designs, from which a single design is ultimately selected using one of the strategies described in Section~\ref{sec:algorithm1}. 
To assess performance, we compute the cost $\widetilde{c}_r$, $r=1, \dots, R$, of each of the $R$ final candidate designs (see Section \ref{sec:problemandnotation}) and summarize them through three statistics: the minimum cost ($\Cmin$), the mean cost ($\Cmean$), and the maximum cost ($\Cmax$). %, defined as
% \begin{equation}
%     \Cmean = 100 \times \frac{1}{R} \sum_{r=1}^{R} {\widetilde{c}_r}, \qquad
%     \Cmin = 100 \times \min_{r=1,\dots,R} {\widetilde{c}_r}, \qquad
%     \Cmax = 100 \times \max_{r=1,\dots,R} \widetilde{c}_r.
% \end{equation}
In order to compare the balance of our proposed algorithm with that of the cube method, we also define the relative values of these statistics by dividing them by the estimated cost of the cube method $\widehat{\widetilde{c}}_{\textup{cube}}$, yielding the percent relative minimum cost ($\RCmin$), the percent relative mean cost ($\RCmean$), and the percent relative maximum cost ($\RCmax$):
\begin{equation}
    \RCmin =  100 \times \frac{\Cmin}{\widehat{\widetilde{c}}_{\textup{cube}}}, \qquad
    \RCmean = 100 \times \frac{\Cmean}{\widehat{\widetilde{c}}_{\textup{cube}}}, \qquad
    \RCmax =  100 \times \frac{\Cmax}{\widehat{\widetilde{c}}_{\textup{cube}}}.
\end{equation}
Note that if the selected design is the best among the $R$ candidates, its relative cost corresponds to the $\RCmin$ statistic.
The results are reported in Table~\ref{tab:simulateddata:relativecost}. 

\begin{table}[htb!]
    \caption{Comparison of Algorithms \ref{algo:yves} and \ref{algo:esther:caren} in terms of cost (percent minimum ($\RCmin$ in \%), mean ($\RCmean$ in \%), and maximum ($\RCmax$ in \%) cost relative to the cost of the cube method) and computation time, for different population sizes $N$ and numbers of balancing variables $p$. \\}
    \label{tab:simulateddata:relativecost}
    \centering
    \begin{tabular}{llrrrrrrrr} 
        \hline
        \multirow{2}{*}{$N$} & \multirow{2}{*}{$p$} 
        & \multicolumn{4}{c}{Algorithm 1} 
        & \multicolumn{4}{c}{Algorithm 2} \\
        \cmidrule(lr){3-6} \cmidrule(lr){7-10}
        & 
        & $\RCmin$ & $\RCmean$ & $\RCmax$ & Time (s) 
        & $\RCmin$ & $\RCmean$ & $\RCmax$ & Time (s) \\
        \hline
        200 & 10 & 13.81 & 16.38 & 21.75 & 7.08 & 16.24 & 16.26 & 16.28 & 7.44 \\
        200 & 15 & 14.84 & 15.16 & 15.34 & 331.96 & 14.36 & 14.45 & 14.48 & 411.44 \\
        500 & 10 & 3.61 & 3.78 & 4.05 & 36.49 & 3.59 & 3.79 & 4.27 & 55.25 \\
        500 & 15 & 16.18 & 16.77 & 17.16 & 1546.91 & 16.22 & 16.32 & 16.40 & 1053.22 \\
        \hline
    \end{tabular}
\end{table}

Overall, both algorithms exhibit similar performance in terms of minimum, mean, and maximum cost across all configurations, indicating comparable behavior in average and variability. 
Across all different settings, the cost achieved by the cube method is reduced by at least a factor of four for all $R = 8$ candidate designs produced by the proposed algorithms.
Specifically, $\RCmean$ is close to 15\% for $N=200, p=10,15$ and $N=500, p=15$, and even less than $4\%$ for $N=200, p=10$.
Turning to $\RCmax$, all values are below 22\%, indicating that even the least well-balanced selected design improves by almost a foctor of five the balance achieved by the cube method.

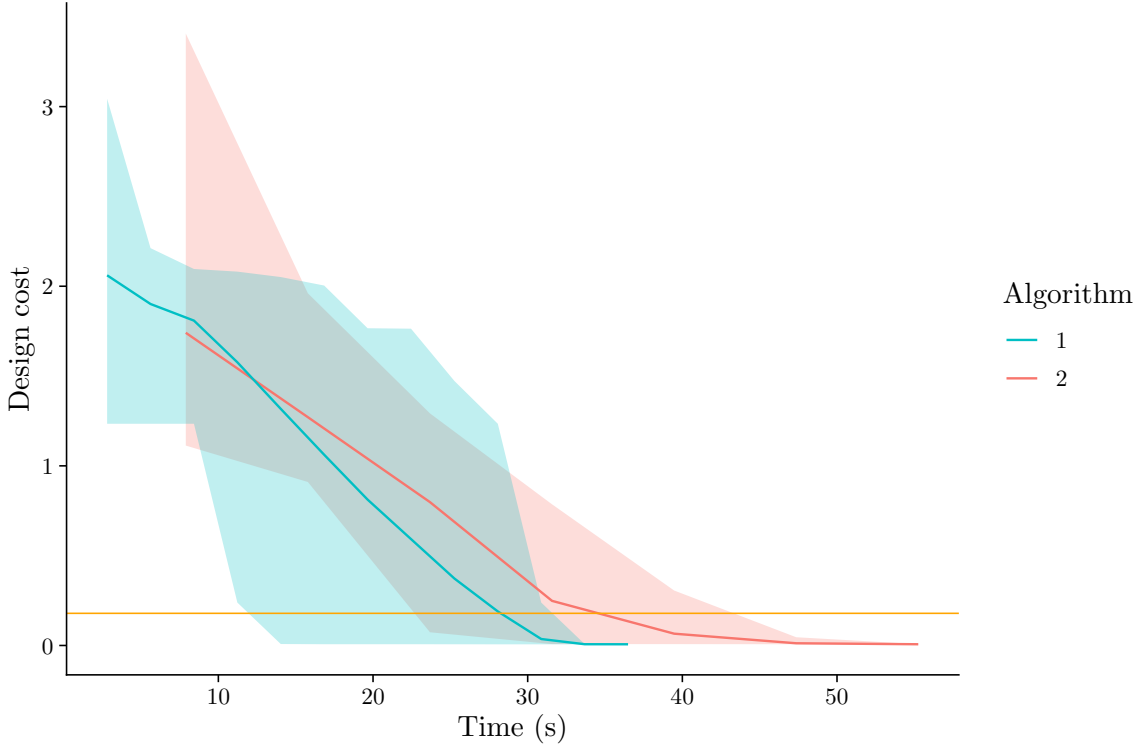
\begin{figure}[htb!]
    \centering
    %\input{figureN=500p=10.tex}
    % Created by tikzDevice version 0.12.6 on 2026-05-05 16:32:48
% !TEX encoding = UTF-8 Unicode
\begin{tikzpicture}[x=1pt,y=1pt]
\definecolor{fillColor}{RGB}{255,255,255}
\path[use as bounding box,fill=fillColor,fill opacity=0.00] (0,0) rectangle (433.62,289.08);
\begin{scope}
\path[clip] (  0.00,  0.00) rectangle (433.62,289.08);
\definecolor{drawColor}{RGB}{255,255,255}
\definecolor{fillColor}{RGB}{255,255,255}

\path[draw=drawColor,line width= 0.6pt,line join=round,line cap=round,fill=fillColor] (  0.00,  0.00) rectangle (433.62,289.08);
\end{scope}
\begin{scope}
\path[clip] ( 27.31, 30.69) rectangle (363.17,283.58);
\definecolor{fillColor}{RGB}{255,255,255}

\path[fill=fillColor] ( 27.31, 30.69) rectangle (363.17,283.58);
\definecolor{fillColor}{RGB}{248,118,109}

\path[fill=fillColor,fill opacity=0.25] ( 72.19,272.08) --
	(118.14,174.31) --
	(164.10,129.10) --
	(210.05, 94.88) --
	(256.00, 62.44) --
	(301.95, 44.85) --
	(347.90, 42.26) --
	(347.90, 42.18) --
	(301.95, 42.21) --
	(256.00, 42.26) --
	(210.05, 42.28) --
	(164.10, 46.72) --
	(118.14,103.29) --
	( 72.19,116.99) --
	cycle;

\path[] ( 72.19,272.08) --
	(118.14,174.31) --
	(164.10,129.10) --
	(210.05, 94.88) --
	(256.00, 62.44) --
	(301.95, 44.85) --
	(347.90, 42.26);

\path[] (347.90, 42.18) --
	(301.95, 42.21) --
	(256.00, 42.26) --
	(210.05, 42.28) --
	(164.10, 46.72) --
	(118.14,103.29) --
	( 72.19,116.99);
\definecolor{fillColor}{RGB}{0,191,196}

\path[fill=fillColor,fill opacity=0.25] ( 42.58,247.55) --
	( 58.92,191.31) --
	( 75.26,183.46) --
	( 91.60,182.47) --
	(107.94,180.43) --
	(124.28,177.23) --
	(140.62,161.16) --
	(156.96,160.99) --
	(173.30,141.31) --
	(189.64,125.20) --
	(205.98, 57.90) --
	(222.32, 42.29) --
	(238.66, 42.24) --
	(238.66, 42.18) --
	(222.32, 42.18) --
	(205.98, 42.18) --
	(189.64, 42.18) --
	(173.30, 42.20) --
	(156.96, 42.22) --
	(140.62, 42.22) --
	(124.28, 42.22) --
	(107.94, 42.29) --
	( 91.60, 57.90) --
	( 75.26,125.20) --
	( 58.92,125.20) --
	( 42.58,125.20) --
	cycle;

\path[] ( 42.58,247.55) --
	( 58.92,191.31) --
	( 75.26,183.46) --
	( 91.60,182.47) --
	(107.94,180.43) --
	(124.28,177.23) --
	(140.62,161.16) --
	(156.96,160.99) --
	(173.30,141.31) --
	(189.64,125.20) --
	(205.98, 57.90) --
	(222.32, 42.29) --
	(238.66, 42.24);

\path[] (238.66, 42.18) --
	(222.32, 42.18) --
	(205.98, 42.18) --
	(189.64, 42.18) --
	(173.30, 42.20) --
	(156.96, 42.22) --
	(140.62, 42.22) --
	(124.28, 42.22) --
	(107.94, 42.29) --
	( 91.60, 57.90) --
	( 75.26,125.20) --
	( 58.92,125.20) --
	( 42.58,125.20);
\definecolor{drawColor}{RGB}{248,118,109}

\path[draw=drawColor,line width= 0.9pt,line join=round] ( 72.19,159.45) --
	(118.14,127.68) --
	(164.10, 95.71) --
	(210.05, 58.54) --
	(256.00, 46.19) --
	(301.95, 42.58) --
	(347.90, 42.21);
\definecolor{drawColor}{RGB}{0,191,196}

\path[draw=drawColor,line width= 0.9pt,line join=round] ( 42.58,181.08) --
	( 58.92,170.28) --
	( 75.26,164.03) --
	( 91.60,148.34) --
	(107.94,130.81) --
	(124.28,113.54) --
	(140.62, 96.66) --
	(156.96, 81.80) --
	(173.30, 66.95) --
	(189.64, 54.56) --
	(205.98, 44.18) --
	(222.32, 42.22) --
	(238.66, 42.20);
\definecolor{drawColor}{RGB}{255,165,0}

\path[draw=drawColor,line width= 0.6pt,line join=round] ( 27.31, 53.85) -- (363.17, 53.85);
\end{scope}
\begin{scope}
\path[clip] (  0.00,  0.00) rectangle (433.62,289.08);
\definecolor{drawColor}{RGB}{0,0,0}

\path[draw=drawColor,line width= 0.6pt,line join=round,line cap=rect] ( 27.31, 30.69) --
	( 27.31,283.58);
\end{scope}
\begin{scope}
\path[clip] (  0.00,  0.00) rectangle (433.62,289.08);
\definecolor{drawColor}{RGB}{0,0,0}

\node[text=drawColor,anchor=base east,inner sep=0pt, outer sep=0pt, scale=  0.88] at ( 22.36, 38.72) {0};

\node[text=drawColor,anchor=base east,inner sep=0pt, outer sep=0pt, scale=  0.88] at ( 22.36,106.33) {1};

\node[text=drawColor,anchor=base east,inner sep=0pt, outer sep=0pt, scale=  0.88] at ( 22.36,173.94) {2};

\node[text=drawColor,anchor=base east,inner sep=0pt, outer sep=0pt, scale=  0.88] at ( 22.36,241.55) {3};
\end{scope}
\begin{scope}
\path[clip] (  0.00,  0.00) rectangle (433.62,289.08);
\definecolor{drawColor}{RGB}{0,0,0}

\path[draw=drawColor,line width= 0.6pt,line join=round] ( 24.56, 41.75) --
	( 27.31, 41.75);

\path[draw=drawColor,line width= 0.6pt,line join=round] ( 24.56,109.36) --
	( 27.31,109.36);

\path[draw=drawColor,line width= 0.6pt,line join=round] ( 24.56,176.97) --
	( 27.31,176.97);

\path[draw=drawColor,line width= 0.6pt,line join=round] ( 24.56,244.58) --
	( 27.31,244.58);
\end{scope}
\begin{scope}
\path[clip] (  0.00,  0.00) rectangle (433.62,289.08);
\definecolor{drawColor}{RGB}{0,0,0}

\path[draw=drawColor,line width= 0.6pt,line join=round,line cap=rect] ( 27.31, 30.69) --
	(363.17, 30.69);
\end{scope}
\begin{scope}
\path[clip] (  0.00,  0.00) rectangle (433.62,289.08);
\definecolor{drawColor}{RGB}{0,0,0}

\path[draw=drawColor,line width= 0.6pt,line join=round] ( 84.45, 27.94) --
	( 84.45, 30.69);

\path[draw=drawColor,line width= 0.6pt,line join=round] (142.67, 27.94) --
	(142.67, 30.69);

\path[draw=drawColor,line width= 0.6pt,line join=round] (200.88, 27.94) --
	(200.88, 30.69);

\path[draw=drawColor,line width= 0.6pt,line join=round] (259.10, 27.94) --
	(259.10, 30.69);

\path[draw=drawColor,line width= 0.6pt,line join=round] (317.31, 27.94) --
	(317.31, 30.69);
\end{scope}
\begin{scope}
\path[clip] (  0.00,  0.00) rectangle (433.62,289.08);
\definecolor{drawColor}{RGB}{0,0,0}

\node[text=drawColor,anchor=base,inner sep=0pt, outer sep=0pt, scale=  0.88] at ( 84.45, 19.68) {10};

\node[text=drawColor,anchor=base,inner sep=0pt, outer sep=0pt, scale=  0.88] at (142.67, 19.68) {20};

\node[text=drawColor,anchor=base,inner sep=0pt, outer sep=0pt, scale=  0.88] at (200.88, 19.68) {30};

\node[text=drawColor,anchor=base,inner sep=0pt, outer sep=0pt, scale=  0.88] at (259.10, 19.68) {40};

\node[text=drawColor,anchor=base,inner sep=0pt, outer sep=0pt, scale=  0.88] at (317.31, 19.68) {50};
\end{scope}
\begin{scope}
\path[clip] (  0.00,  0.00) rectangle (433.62,289.08);
\definecolor{drawColor}{RGB}{0,0,0}

\node[text=drawColor,anchor=base,inner sep=0pt, outer sep=0pt, scale=  1.10] at (195.24,  7.64) {Time (s)};
\end{scope}
\begin{scope}
\path[clip] (  0.00,  0.00) rectangle (433.62,289.08);
\definecolor{drawColor}{RGB}{0,0,0}

\node[text=drawColor,rotate= 90.00,anchor=base,inner sep=0pt, outer sep=0pt, scale=  1.10] at ( 13.08,157.13) {Design cost};
\end{scope}
\begin{scope}
\path[clip] (  0.00,  0.00) rectangle (433.62,289.08);
\definecolor{fillColor}{RGB}{255,255,255}

\path[fill=fillColor] (374.17,129.57) rectangle (428.12,184.69);
\end{scope}
\begin{scope}
\path[clip] (  0.00,  0.00) rectangle (433.62,289.08);
\definecolor{drawColor}{RGB}{0,0,0}

\node[text=drawColor,anchor=base west,inner sep=0pt, outer sep=0pt, scale=  1.10] at (379.67,170.55) {Algorithm};
\end{scope}
\begin{scope}
\path[clip] (  0.00,  0.00) rectangle (433.62,289.08);
\definecolor{fillColor}{RGB}{255,255,255}

\path[fill=fillColor] (379.67,149.53) rectangle (394.12,163.98);
\definecolor{drawColor}{RGB}{0,191,196}

\path[fill=fillColor,fill opacity=0.25] (380.38,150.24) rectangle (393.41,163.27);
\definecolor{drawColor}{RGB}{0,191,196}

\path[draw=drawColor,line width= 0.9pt,line join=round] (381.11,156.75) -- (392.68,156.75);
\end{scope}
\begin{scope}
\path[clip] (  0.00,  0.00) rectangle (433.62,289.08);
\definecolor{fillColor}{RGB}{255,255,255}

\path[fill=fillColor] (379.67,135.07) rectangle (394.12,149.53);
\definecolor{drawColor}{RGB}{248,118,109}

\path[fill=fillColor,fill opacity=0.25] (380.38,135.78) rectangle (393.41,148.81);
\definecolor{drawColor}{RGB}{248,118,109}

\path[draw=drawColor,line width= 0.9pt,line join=round] (381.11,142.30) -- (392.68,142.30);
\end{scope}
\begin{scope}
\path[clip] (  0.00,  0.00) rectangle (433.62,289.08);
\definecolor{drawColor}{RGB}{0,0,0}

\node[text=drawColor,anchor=base west,inner sep=0pt, outer sep=0pt, scale=  0.88] at (399.62,153.72) {1};
\end{scope}
\begin{scope}
\path[clip] (  0.00,  0.00) rectangle (433.62,289.08);
\definecolor{drawColor}{RGB}{0,0,0}

\node[text=drawColor,anchor=base west,inner sep=0pt, outer sep=0pt, scale=  0.88] at (399.62,139.27) {2};
\end{scope}
\end{tikzpicture}
    \caption{Evolution of the design cost across iterations for $N=500$, $p=10$. 
    The blue line (respectively red line) represents the mean cost over the $R=8$ designs at each iteration, with the corresponding minimum and maximum values shown as a shaded ribbon for Algorithm~\ref{algo:yves} (respectively Algorithm~\ref{algo:esther:caren}). 
    The orange line corresponds to the estimated cost of the cube method.}
    \label{fig:convergence}
\end{figure}

Regarding computational efficiency, the running times increase with both the population size $N$ and the number of balancing variables $p$. 
For instance, for $N=200, p=10$, the running time is 7 seconds, whereas it rises to 1546 seconds in the worst-case scenario $N=500, p=15$.
No method outperforms the others in terms of computation time. 
An example of the evolution of the cost across the iterations of the proposed algorithms is provided in Figure~\ref{fig:convergence}. It shows that both algorithms follow a similar convergence pattern.

In summary, our proposed algorithms yield designs with a lower cost than the cube method across all configurations.
However, the running time to obtain the final $R = 8$ well-balanced designs increases substantially with the population size and the number of balancing variables.

\subsection{Real data}

We also evaluate the proposed algorithms on a real dataset. 
We consider the dataset MU284 that contains information about $N=284$ municipalities \citep{sar:swe:wre:92} and is available in the R package sampling \citep{til:mat:25}.
The inclusion probabilities $\pib = (\pi_1, \dots, \pi_N)$ are set proportional to variable $P75$, the population in 1975.
Moreover, different expected sample sizes $n$ are considered: $10, 20, 50,$ and $100$, and the inclusion probabilities are computed such that $\sum_{k=1}^N \pi_k = n$.

The goal is to obtain a sampling design satisfying the inclusion probabilities~$\pib$, while achieving the best possible balance on a set of $p=6$ auxiliary variables: $P75$, $CS82$, $SS82$, $S82$, $ME84$, $REV84$ (see \citet{sar:swe:wre:92} for a full description of the variables).
For each sample size ($n=10, 20, 50, 100$), we generate $R$ well-balanced designs using Algorithms~\ref{algo:yves} and~\ref{algo:esther:caren} under the same settings as those used for the simulated data (see Section~\ref{sec:simu:simulateddata}).
Both algorithms are evaluated using the same criteria as in Section~\ref{sec:simu:simulateddata}.
Results are reported in Table~\ref{tab:realdata:relativecost}. 

\begin{table}[htb!]
    \caption{Comparison of Algorithms~\ref{algo:yves} and~\ref{algo:esther:caren} in terms of cost (minimum ($\RCmin$), mean ($\RCmean$), and maximum ($\RCmax$) relative to the cube method, in \%) and computation time, for different sample sizes $n$, using the MU284 dataset. \\ }
    \label{tab:realdata:relativecost}
    \centering
        \begin{tabular}{llrrrrrrrr}
            \hline
            \multirow{2}{*}{$n$} & \multirow{2}{*}{$p$} 
            & \multicolumn{4}{c}{Algorithm 1} 
            & \multicolumn{4}{c}{Algorithm 2} \\
            \cmidrule(lr){3-6} \cmidrule(lr){7-10}
            & 
            & $\RCmin$ & $\RCmean$ & $\RCmax$ & Time (s) 
            & $\RCmin$ & $\RCmean$ & $\RCmax$ & Time (s) \\
            \hline
            10  & 7  & 24.28 & 26.57 & 27.40 & 638.40  & 29.69 & 29.86 & 30.03 & 731.69 \\
            20  & 7  & 24.64 & 25.15 & 25.72 & 703.75  & 33.53 & 33.78 & 33.94 & 1181.99 \\
            50  & 7  & 55.06 & 55.50 & 56.11 & 1094.86 & 30.59 & 30.71 & 30.82 & 1824.43 \\
            100 & 7  & 4.42  & 4.81  & 5.17  & 1067.56 & 4.14  & 4.20  & 4.29  & 2549.64 \\
            \hline
        \end{tabular}
\end{table}

Algorithm~\ref{algo:yves} and~\ref{algo:esther:caren} exhibit similar performance in terms of cost for $n = 10$ and $20$. Across the $R = 8$ final designs, the cost relative to the cube method ranges between $24\%$ and $34\%$.
For $n=50$, all $R = 8$ final designs generated by Algorithm~\ref{algo:yves} and~\ref{algo:esther:caren} have relative costs of approximately $55\%$ and $30\%$, respectively.
For $n=100$, the corresponding values fall to approximately 4 to $5\%$ for both Algorithms~\ref{algo:yves} and~\ref{algo:esther:caren}.
Across all values of $n$, the cost of the cube method is reduced for all $R = 8$ final designs produced by the proposed algorithms.

Regarding computational efficiency, the running times increase with the sample size $n$. 
For instance, for $n=10$, the running times of the algorithms are 638 and 731 seconds, and it rises to 1067 and 2550 seconds in the worst-case scenario $n=100$.
The two algorithms show similar computation times, with neither clearly outperforming the other.

In summary, our proposed algorithms yield designs with a lower cost than the cube method across all values of $n$.
However, the time required to obtain the final $R = 8$ well-balanced designs is higher than in the simulated data setting (Section~\ref{sec:simu:simulateddata}) and increases substantially with the sample size.
This large observed running time can be explained by the asymmetry of the variable used to construct the inclusion probabilities, which results in some inclusion probabilities being equal to one.

\section{Conclusion}\label{sec:conclusion}

This article presents a novel approach for constructing balanced sampling designs using a genetic algorithm. This approach iteratively improves the balance of a set of candidate sampling designs. While we cannot guarantee identification of the globally optimal balanced design, the improvement in balance can be substantial compared to the cube method. Our proposed approach generates sampling designs close to the optimal design without having to list all possible samples, while exactly respecting the inclusion probabilities. Two alternative algorithms are presented, only one of which can be adapted to parallel computing. This represents an advantage in the context of large populations. We illustrate the behavior of both proposed algorithms via simulation studies on simulated and real data. The results of these studies confirm that the generated sampling designs outperform the cube method in terms of design cost. Our proposed method has the potential to improve the balance in practical applications in both sampling and experimental designs.

\bibliographystyle{apalike}
%\bibliography{bibyves}

\clearpage
\appendix

\section{Algorithm for systematic sample selection with unequal inclusion probabilities} \label{appendix:systematicsample}

\begin{algorithm}[htb!]
\caption{Algorithm for systematic sample selection with unequal inclusion probabilities}\label{algo:systematic:sample}
\begin{algorithmic}[1]
\REQUIRE Inclusion probabilities $\pi_1,\dots,\pi_N$ such that $\sum_{k \in U} \pi_k = n$; optional start $u\in[0,1]$
\ENSURE Selected indices $k_1,\dots,k_n$
\STATE Set $V_0 \leftarrow 0$
\FOR{$k=1$ \TO $N$}
  \STATE $V_k \leftarrow V_{k-1} + \pi_k$
\ENDFOR
\IF{$u$ is not specified}
  \STATE Draw $u \sim \mathrm{Unif}(0,1)$
\ENDIF
\STATE Initialize $j \leftarrow 0$, $k \leftarrow 1$
\FOR{$j=0$ \TO $n-1$}
  \STATE $t \leftarrow u + j$
  \WHILE{$t > V_k$}
    \STATE $k \leftarrow k + 1$
  \ENDWHILE
  \STATE Select unit $k_{j+1} \leftarrow k$ \COMMENT{$V_{k-1} < t \le V_k$}
\ENDFOR
\RETURN $\{k_1,\dots,k_n\}$
\end{algorithmic}
\end{algorithm}

\newpage 
\section{Algorithm for systematic design with unequal inclusion probabilities} \label{appendix:systematicdesign}

\begin{algorithm}[htb!]
\caption{Algorithm for systematic design with unequal inclusion probabilities}\label{algo:systematic:design}
\begin{algorithmic}[1]
\REQUIRE Inclusion probabilities $\pi_1,\dots,\pi_N$ with $\sum_{k \in U}\pi_k = n$
\ENSURE Matrix of samples $\Sb$ and probabilities $\pb$

\STATE \textbf{Compute cumulative sums}
\STATE $V_0 \leftarrow 0$
\FOR{$k=1$ \TO $N-1$}
    \STATE $V_k \leftarrow V_{k-1} + \pi_k$
\ENDFOR

\STATE \textbf{Compute fractional parts}
\FOR{$k=0$ \TO $N-1$}
    \STATE $v_k \leftarrow V_k \bmod 1$
\ENDFOR

\STATE \textbf{Sort fractional parts}
\STATE Obtain ordered values $v_{(0)} \le \dots \le v_{(N-1)}$
\STATE Set $v_{(N)} \leftarrow 1$

\STATE \textbf{Compute interval probabilities and midpoints}
\FOR{$k=1$ \TO $N$}
    \STATE $\displaystyle p_k \leftarrow v_{(k+1)} - v_{(k)}$\\[1mm]
    \STATE $\displaystyle r_k \leftarrow \frac{v_{(k)} + v_{(k+1)}}{2}$
\ENDFOR

\STATE \textbf{Generate systematic samples}
\STATE Initialize empty matrix $\Sb$ and empty vector $\pb$
\FOR{$k=0$ \TO $N$}
    \IF{$p_k > 0$}
        \STATE $\mathbf{s}_k \leftarrow$ systematic sample with $(\pi_1,\dots,\pi_N; r_k)$
        \STATE Add $\mathbf{s}_k$ as a column of $\mathbf{S}$
        \STATE Add $p_k$ as an element of $\pb$
    \ENDIF
\ENDFOR

\RETURN $\Sb, \pb$
\end{algorithmic}
\end{algorithm}

\end{document}